\documentclass[aps,amssymb]{revtex4}

\usepackage{graphicx}

\begin{document}
\renewcommand{\thefootnote}{\fnsymbol{footnote}} 
\renewcommand{\theequation}{\arabic{section}.\arabic{equation}}

\title{Adsorption of polyelectrolytes from semi-dilute solutions\\ on an oppositely charged surface}

\author{Manoel Manghi,$^{a,b}$ and Miguel Aubouy$^b$}

\affiliation{$^a$ Laboratoire de Physique Th\'eorique, Universit\'e de
Toulouse, CNRS, 31062 Toulouse, France\\
$^b$ SPrAM, Universit\'e Joseph Fourier, CNRS, CEA-Grenoble, 38054 Grenoble, France}

\date{\today}

\begin{abstract}
\noindent We propose a detailed description of the structure of the layer formed by
polyelectrolyte chains adsorbed onto an oppositely charged surface in the
semi-dilute regime. We combine the mean-field Poisson-Boltzmann-Edwards theory and the scaling functional theory to describe the variations of the monomer concentration, the electrostatic potential, and the local grafting density with the distance to the surface. For long polymers, we find that the effective charge of the decorated surface (surface \textit{plus} adsorbed polyelectrolytes) can be much larger than the bare charge of the surface at low salt concentration, thus providing an experimental route to a "supercharging" type of effect.
\end{abstract}

\maketitle

\section{Introduction}


When a charged solid surface (charge density $\sigma_{0}$) is
exposed to a solution of polyelectrolytes carrying opposite ions
($Z$ charges per chain), a counter-ion exchange occurs such that
the chains adsorb and the counter-ions initially retained captive
by electrostatic attraction are released in the solution. This is
because the system gains the translational entropy of the
micro-ions, which is $Zk_BT$ (where $k_BT$ is the thermal
energy), and only looses $k_{B}T$ for each adsorbed chain. Then at
sufficiently high charge density, adsorbed polyelectrolytes form a
dense layer of coils strongly bounded to the solid. In some
favorable cases, because the macro-ions have very long tails
dandling in the solution, the adsorbed layer alone may carry a
surface charge density larger than $\sigma_{0}$. When the solution
has been rinsed away, the decorated surface (solid surface
\textit{plus} attached chains) then behaves as a new charged
system surrounded by counter-ions. This feature is the basic
principle of a new device to build controlled charged multi-layers
of alternated sign~\cite{Decher}.

On the theoretical side, the structure of the adsorbed layer is known in the
dilute regime, where the chains in the bulk are isolated coils, and the
solution does not affect much the
layer~\cite{Joanny,DobryninDeshkovski,Dobrynin,Dobrynin2,varoqui,stoll1,stoll2}. Comparatively, much
less is known about adsorption from \textit{semi-dilute solutions}. Because the
dilute regime for polyelectrolytes is found at vanishing monomer concentrations
($\phi_{b}^{*}\sim N^{-2}$ where $N$ is the index of polymerization), the
semi-dilute regime is certainly relevant from an experimental point of view.
Note that the Decher process takes place in the semi-dilute regime in
Ref.~\cite{Decher2}. Borukhov \textit{et al.} have numerically solved the
non-linear mean-field system of equations in the semi-dilute regime for
repulsive surfaces~\cite{Borukhov}. They also used scaling arguments
to determine the relevant characteristic length scale in terms of the total electrostatic
potential drop, but no detailed analytical description of the layer is done. Besides,
the case of attractive surfaces does not simply follow from the repulsive case
explicitly treated by Borukhov \textit{et al.} since the concentration profiles
are significantly affected by the boundary conditions. Ch\^{a}tellier and Joanny
have solved the linearized mean-field system of equations suitable to describe
how the charged surface affects the semi-dilute solution~\cite{Chatellier}. However, this approach is restricted to small perturbations. By solving numerically the self-consistent field equations using the ground state dominance approximation, Wang explored the influence of various parameters (surface charge, salt concentration, bulk concentration) on the occurrence of charge inversion~\cite{wang}. The latter was found to be strong at high salt concentration. 

In any case, these works do not provide the extension, $H$, of the
adsorbed layer (which does not identify with the characteristic
relaxation length of the concentration), and the amount of
material, $\Gamma$, attached to the solid surface (which does not
identifies with the excess material driven to the interface
$\int(\phi-\phi_b)\,dz$, where $\phi$ is the monomer concentration, as defined
in Refs.~\cite{Borukhov,Andelman,wang}). There is a technical reason for that~: the
mean-field approach for adsorbed polyelectrolytes in semi-dilute conditions has
to be supplemented by another approach to provide the full picture. This is
because the variations of the polymeric concentration and the electrostatic
potential does not carry enough information to completely describe the layer.
This matters in practical situation since $\Gamma$ is directly related
to the charge density carried by the decorated surface after
removal from the solution. Accordingly, this quantity deserves a
special attention. Moreover, it has been seen experimentally that the extension
and the amount of material of the adsorbed layer vary significantly with
$N$~\cite{Chibowski}. This effect is not explained using the mean-field theory
in the ground-state dominance approximation~\cite{PGGbook} alone, which is used
in Refs.~\cite{Borukhov,Chatellier,Andelman,wang}. Indeed within this approach, the
semi-dilute solution is described with $\phi$ and the adsorbed chains cannot be distinguished from free chains in the semi-dilute layer.

In this article, we propose a complete description of the adsorbed
layer in equilibrium with the semi-dilute solution. In particular,
we estimate $\Gamma$ and $H$ as a function of $N$, $\sigma_0$
and $\phi_b$. We use a mean-field approach supplemented by a
scaling type of description. The mean-field theory that we
consider is the celebrated Poisson-Boltzmann-Edwards' (PBE)
description of the polyelectrolyte solution (see the
reviews~\cite{Andelman,Barrat}). The scaling approach that we use
is the Scaling Functional Theory (SFT)~\cite{Manomacromol} that we adapt to treat
semi-dilute solutions of polyelectrolytes.

We shall consider a rather standard situation: long ($N\gg 1$), linear,
fully flexible (the persistence length is identified with the monomer size
$a$), polymer chains, small fraction of charges ($f\ll 1$), in $\Theta$ solvent
conditions ($v=0$, and the monomer/monomer interaction is dominated by three
body excluded volume interactions), bare surface charge $\sigma_0$. We assume no added salt and no specific interaction with the surface, as a first hint into the full problem. For definiteness, we suppose (without loss of generality) that the surface is
negatively charged with a bare surface charge density $-\sigma_{0} $ (with
$\sigma_0>0$), and the chain carries $Z=fN$ quenched elementary positive
charges $+e$ where $f$ is the fraction of charged monomers.

\begin{figure}[ht]
\begin{center}
\includegraphics[height=7cm]{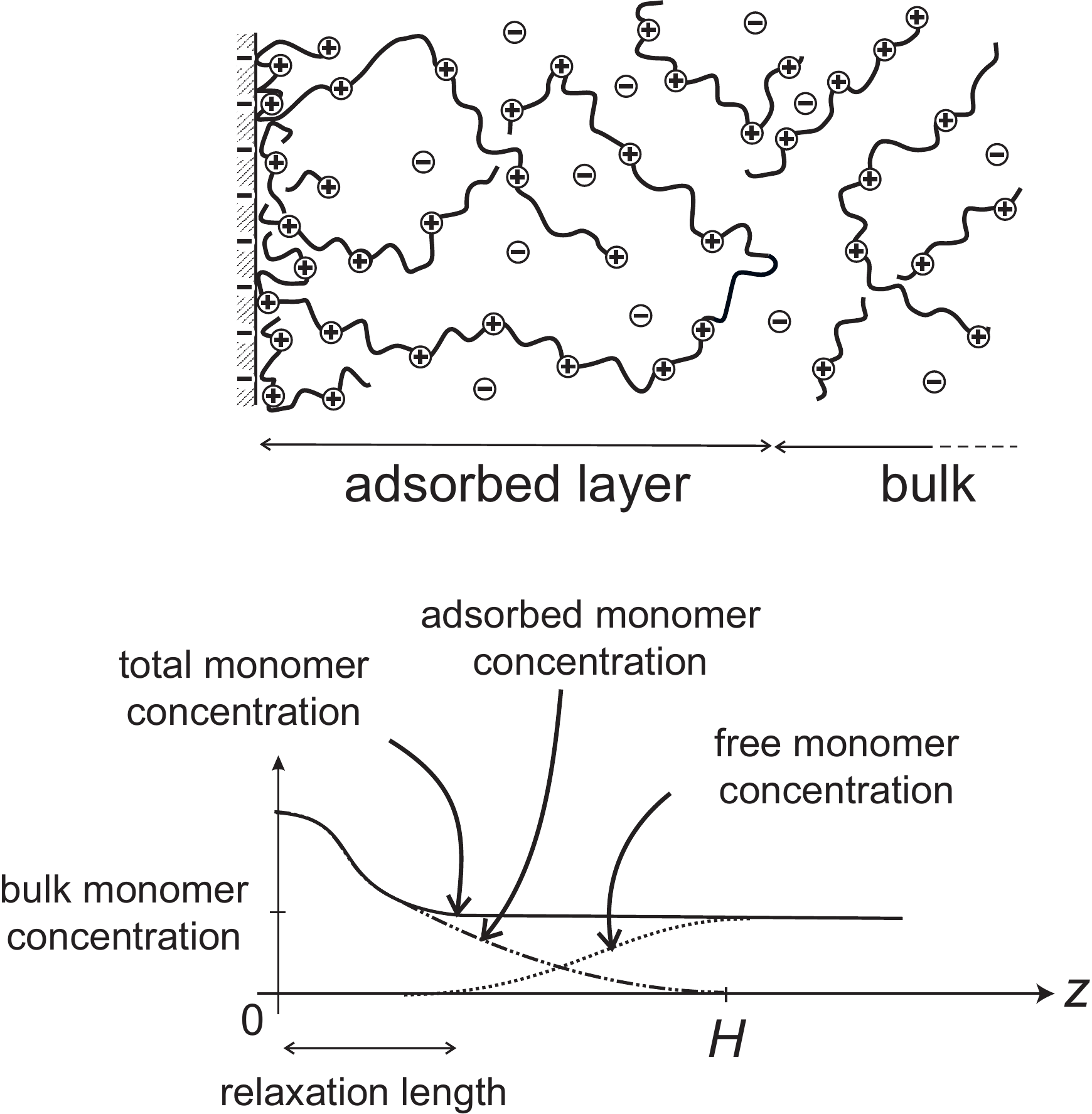}
\end{center}
\caption{Sketch of the adsorbed polyelectrolyte layer
exposed to a semi-dilute solution. Some long dandling tails and
loops are protruding in the bulk. The total monomer concentration
profile is thus the sum of the adsorbed monomer concentration and
the free monomer concentration. We note that the relaxation length
of the monomer concentration close to the surface is different
from the total adsorbed layer thickness, $H$.}
\label{fig1}
\end{figure}

\begin{figure}[ht]
\begin{center}
\includegraphics[height=7cm]{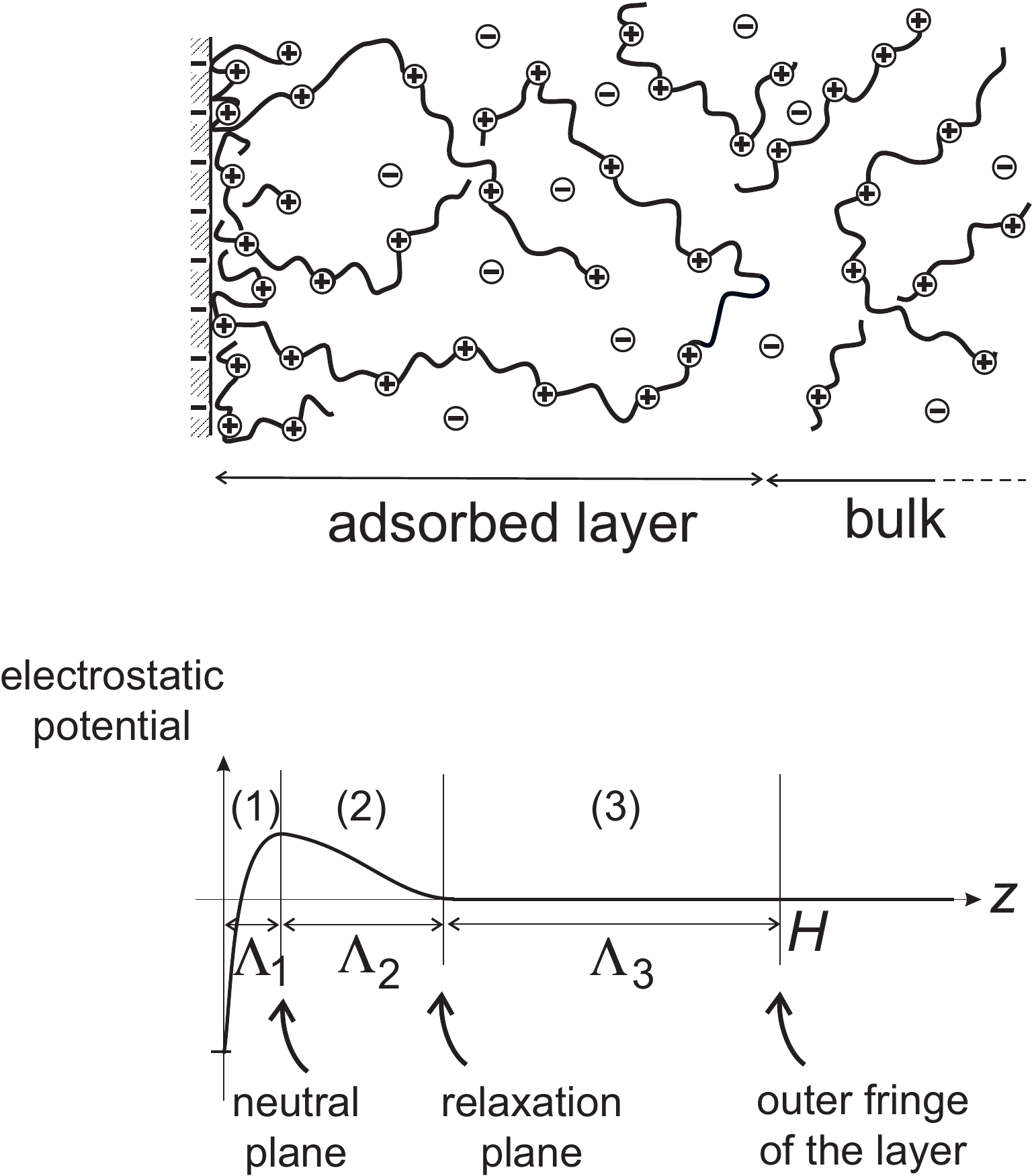}
\end{center}
\caption{Sketch of the variations of the mean electrostatic
potential in the direction perpendicular to the surface. It starts with a
negative value at the negatively charged surface, increases until it reaches
a maximum at the "neutral plane", where the electric field vanishes. Then it
relaxes until it reaches its average value in the bulk, 0, at the relaxation
plane and is constant in the outer region of the adsorbed layer.}
\label{fig2}
\end{figure}

In view of the number of parameters, this issue is a formidable task. To
proceed further, our analysis is based on two major assumptions~:\\
1. That we may find long polyelectrolyte tails belonging to
adsorbed chains far from the surface (protruding in the solution), will be
our first assumption (see Fig.~\ref{fig1}). This hypothesis is inspired by the
results found with long neutral chains adsorbed from semi-dilute solution,
where we know that the concentration field decreases over a distance
comparable to the bulk blob size $\xi_{b}\sim \phi_{b}^{-3/4}$ (the
characteristic length scale of the solution), whereas the layer of adsorbed
chains extend over much larger distances $H\cong R_{G}\sim
N^{1/2}\phi_{b}^{7/8}$~\cite{Manomacromol,Marques,Daoud}. In this example,
the essential idea is that the collective response of the chains screens the
perturbation induced by the surface at a characteristic length scale which
does not depend on the chain size~\cite{PGG81}. A similar feature is found
with polyelectrolytes: in the semi-dilute regime, the Debye-H\"{u}ckel
length of the solution, $\kappa^{-1}$, is independent of the chain
length. One may possibly argue that polyelectrolytes lie quite flat on
the surface, as with dilute solutions. This is however very unlikely since for such
picture extrapolated to very high concentrations of polyelectrolyte, it would
mean that the adsorbed chains adopt an almost two-dimensional configuration.
Clearly, the system of adsorbed chains would gain large amounts of
conformational entropy by allowing more chains to adsorb and recovering an
almost Gaussian structure (thus leaving some tails dandling in the
solution), without perturbing much the distribution of charges in the
vicinity of the solid. In other words, since $\kappa ^{-1}\ll R_{G}\sim
N^{1/2}$ in the limit $N\gg 1$, we expect that the influence of the surface
will be small for the chain structure in equilibrium.\\
2. Looking at the variations of the electrostatic potential, $\Psi(z)$ with
the distance to the surface, $z$, in the dilute regime, we anticipate a very
simple form for $\Psi(z)$ in the semi-dilute regime (see Fig.~\ref{fig2}). From the
mean-field description of the layer in the dilute
regime~\cite{Joanny,DobryninDeshkovski}, we learn that the electrostatic
potential first increases from negative values, then saturates, and eventually
relaxes to zero. That there is a regime of parameters in the semi-dilute regime
where such profile for $\Psi$ will persist to some extent be our second
assumption. This assumption will be justified by the calculation but one
can find a simple argument for it. The monomer concentration profile in the
dilute regime tells us that the osmotic pressure decreases from the solid
surface. Suppose that we increase the concentration of the solution from the
dilute regime into the semi-dilute regime. Because the osmotic pressure close to
the solid surface is much higher than what is found in the solution, we expect
that the presence of the solution will affect the structure of the outer fringe
of the layer, not the structure of the inner regions. Accordingly, we do not
expect that the form of the electrostatic potential will be profoundly affected,
at least in the semi-dilute regime close to the dilute regime.

Following our second hypothesis, we define the plane of vanishing
electrostatic field (such that $d\Psi/dz=0$) as the neutral plane.
Similarly, we set the plane where the electrostatic potential vanishes (such
that $\Psi\cong 0$) as the relaxation plane (we implicit assume that we
haven chosen $\Psi(\infty)=0$). Then the layer might be divided into three
main regions (labeled (1), (2) and (3), see Fig.~\ref{fig2}) such that:
\begin{enumerate}
\item The compensation region stands between the solid surface and the
neutral plane. By definition, the charges provided by the polyelectrolyte
material situated in this region exactly compensate the surface charge.
\item The intermediate region stands between the neutral plane and the
relaxation plane.
\item The outer region stands between the relaxation plane and the outer
border of the layer, as defined by the longest polymeric coil in direct
contact with the surface.
\end{enumerate}
It is important to realize that this distinction, although
\textit{a priori}, is not arbitrary. Each of these region is well
defined from a physical point of view.

In what follows, we first introduce the Scaling Functional Theory
suitable to describe polyelectrolyte semi-dilute layers
(Section~2). Then we propose a detailed picture for the adsorbed
layer in Section~3. Section~4 presents our concluding remarks focusing on the issue of charge inversion. Scaling laws in polyelectrolyte solutions and the mean-field approach are briefly introduced in Appendices~1 and~2 respectively.


\section{Scaling functional theory}


This theory was first introduced to describe neutral polydisperse polymeric
layers by generalizing the Alexander-de~Gennes model~for brushes~\cite{AGR}.
The basic principle is to describe the layer of adsorbed chains as a
thermodynamical ensemble of tails and loops where the entropy associated
with the polydispersity in size compete with the elastic energy and the
monomer excluded volume. From a formal point of view, the SFT approach is a
functional theory where the Hamiltonian is simplified according as
follows~\cite{ManoPRE}: \textit{a)} loops are cut into pseudo-tails, \textit{b)}
the tails and pseudo-tails are described by a single function $z$, where $z(n)$
is the position of the \textit{n}th monomer, $n$, of each tail. This idea proved
successful in describing various situation ranging from polymer brushes,
reversible adsorption, irreversible adsorption, whatever the solvent conditions.

Our aim in this article is not to generalize the SFT to describe
polyelectrolyte chains. Rather, we take advantage of the fact that in the
semi-dilute regime, the chain is both Gaussian and neutral at large scales,
and the system of polymers is thus analogous to a neutral solution provided
that the blob is renormalized according to the physics at small scales.
Essentially, our approach consists in integrating the degrees of freedom of
the various species (including electrostatics interactions) into a proper
description of the blob.

Within the SFT, the system of loops and tails attached to the solid surface
is described by three functions: $S$, the local grafting density of tails
and pseudo-tails (a function of the curvilinear distance $n$), $z$, the
path, and $\phi$, the volume fraction of monomers. These functions are
linked through the conservation of monomers:
\begin{equation}
\phi (z)=\frac{S(n(z))}{\dot{z}(n(z))}  \label{phi}
\end{equation}
For polyelectrolyte semi-dilute layers, the effective free-energy (per
$cm^{2}$) of the system write ($\beta^{-1}=k_{B}T$):
\begin{equation}
\beta F \cong \int_0^N dn \left\{\frac{\left(a^2S(n)\right)^{3/2}}{\xi_e^{3/2}
\dot{z}^{1/2}(n)}+a^{-1}\xi_e^{1/2}S^{3/2}(n)\dot{z}^{3/2}(n) 
-S^{\prime}(n)\ln\left(-\frac{S^{\prime}(n)}{S_{0}}\right) \right\}
\label{SFT energy'} 
\end{equation} 
where $S_0$ is the ``grafting density'' at the surface
and $\xi_e\simeq a(f^{2}\ell_{B})^{-1/3}$ is the electrostatic blob size (in
$\Theta$ solvent conditions, $\ell_B$ is the Bjerrum length in units of $a$, cf.
Appendix~1). The effective free-energy eq.~(\ref{SFT energy'}) is the sum of
three contributions which account respectively for~:\\
\textit{a)} monomer excluded volume interactions. This contribution is $\int
\Pi dz$, where the polymeric contribution to osmotic pressure, $\Pi$,
scales as $k_{B}T/\xi ^{3}$:
\begin{equation}
\Pi \cong k_{B}T\frac{\phi ^{3/2}}{a^{3/2}\xi _{e}^{3/2}}
\end{equation}
With eq.~(\ref{phi}), it is simple to show that
\begin{equation}
\int_{0}^{H}\Pi dz\cong k_BT\int_0^N
\frac{(a^{2}S(n))^{3/2}}{\xi_e^{3/2}\dot{z}^{1/2}(n)}\,dn
\end{equation}
where $H=z(N)$ is the position of the last monomer and is identified as
the layer thickness.\\
\textit{b)} the elasticity of the tail at a scale larger than the blob size (after integration by parts)
\begin{equation}
\beta F_{el} = -\int_0^NdnS^{\prime}(n)\int_0^n\left( \frac{dz/\xi}{dm/g}
\right)^2\frac{dm}{g} \cong a^{-1}\int_0^N\xi_e^{1/2}S^{3/2}(n)
\dot{z}^{3/2}(n)dn
\end{equation}
\textit{c)} the entropy associated to the polydispersity in size of loops
and tails which is similar for charged and neutral monomers as soon as
polymers are flexible~\cite{AGR}.

Minimizing the free-energy eq.~(\ref{SFT energy'}) with respect to $S$ and
$\dot{z}$ yields two equations
\begin{eqnarray}
\frac{3}{2}a^{3}\xi_e^{-3/2} \left(\frac{S}{\dot{z}}\right)^{1/2} + \frac{3}{2a}
\xi_e^{1/2}S^{1/2}\dot{z}^{3/2} + \frac{S^{\prime\prime}} {S^{\prime}} &=& \mu_b
\label{SFT1} \\ \frac12 \left(\frac{a}{\xi_e}\right)^{3/2}
\left(\frac{S}{\dot{z}} \right)^{3/2} - \frac{3}{2a}
\xi_e^{1/2}S^{3/2}\dot{z}^{1/2} &=&\Pi _{b} \label{SFT2}
\end{eqnarray}
These equations are formally similar to what we find in the neutral case
(and their interpretation in physical terms is thus identical), except that
the small scale structure introduces different scaling relationship.
Equation~(\ref{SFT2}) describes the local balance of forces (per unit
surface parallel to the solid): elasticity (first term of lhs) competes with
osmotic pressure (second term of lhs) and the bulk pressure (rhs). Note that
the osmotic pressure of counter-ions is the same in the outer region and in
the bulk so that it does not appear in eq.~(\ref{SFT2}). Equation~(\ref{SFT1})
describes the conservation of the generalized chemical potential.
In equation~(\ref{SFT1}), $\mu_b=\partial\Pi_b/\partial\phi_b$ is the chemical potential in the bulk and scales as $g_b^{-1}$:
\begin{equation}
\mu_b\cong \phi_b^{1/2}\left(\frac{a}{\xi_e}\right)^{3/2}
\end{equation}
The bulk osmotic pressure, $\Pi_b\sim 1/\xi_b^3$, in equation~(\ref{SFT2})
accounts for the pressure induced by the solution.

We emphasis that the SFT system of equations~(\ref{SFT1})--(\ref{SFT2})
assumes that the properties of the tails and loops are that of neutral
strings of blobs, and may only apply to situations where the electrostatics
interactions are screened at distances larger than the mesh size $\xi$ and
especially very close to the substrate.


\section{Structure of the layer}


\subsection{Compensation region (1).}

This region, close to the solid surface, is such that the charges provided
by the polyelectrolyte material exactly compensate the surface charge. Hence
\begin{equation}
\Gamma_1 \cong \sigma_0/f  \label{gamma1}
\end{equation}
and $\frac{d\Psi}{dz}\geq 0$ in this region (at the outer boundary of
the compensation region: $\frac{d\Psi}{dz}=0$). For this reason, we expect
that the counter-ions from polyelectrolytes are not present, $\rho^- \cong 0$. Moreover, at the low salt limit, positive counter-ions (from the charged surface) are negligible since their concentration is proportional to surface/volume and vanishes in the thermodynamic limit.

\begin{figure}[ht]
\begin{center}
\includegraphics[height=3cm]{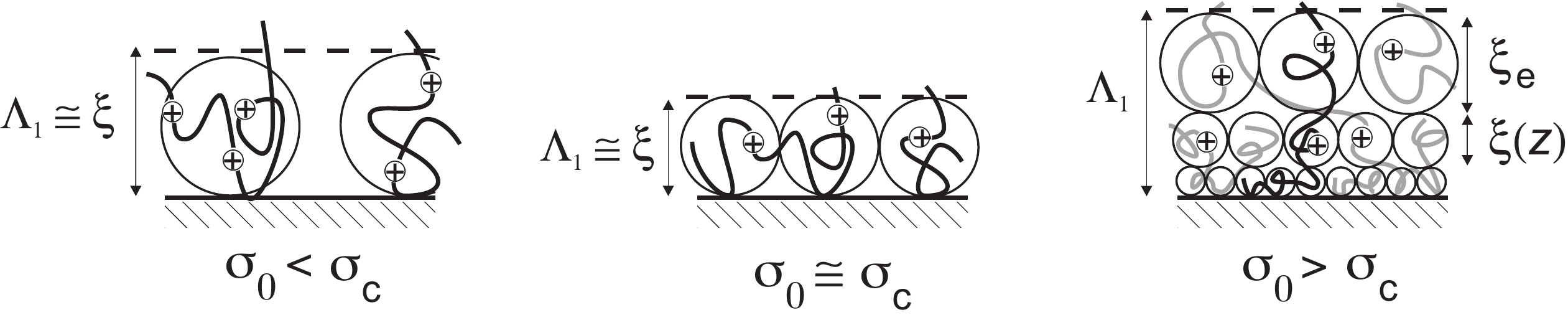}
\end{center}
\caption{Sketch of the compensation region~(1) in the 3 cases:
$\sigma_0<\sigma_c$, $\sigma_0\simeq\sigma_c$ and $\sigma_0>\sigma_c$. The
thickness of this region, $\Lambda_1$ decreases with $\sigma_0$ for $\sigma_0<\sigma_c$ and increases for $\sigma_0>\sigma_c$.}
\label{fig3}
\end{figure}

Each electrostatic blob carries $fg_e$ charges and occupies a surface
$\xi_{e}^{2}$. These relation defines a critical value for $\sigma_{0}$ such
that a single ``carpet" of close packing electrostatic blob is able to fully
compensate the surface charge:
\begin{equation}
\sigma_{c}\cong fa^{-2}  \label{sigmac}
\end{equation}
The monomer volume fraction inside this region is thus $\phi_e$, given by
Eq.~(\ref{phie}). Moreover, the electrostatic energy of a blob interacting
with the surface is of order $k_BT$ for the critical value, $\sigma_0\simeq
\sigma_c$, i.e. of the same order of the electrostatic repulsion between
monomers. Accordingly, we shall distinguish two cases: $\sigma_{0}<\sigma_{c}$
and $\sigma_{0}>\sigma_{c}$ (Fig.~\ref{fig3}).\\

\noindent{\bf III.A.1 Case $\sigma_0<\sigma_c$.} 

In this regime, we rely on the results found by Dobrynin
\textit{et al.}~\cite{DobryninDeshkovski} for the case of dilute
solutions. The electrostatic attraction of charged monomers to the
charged surface is weaker than the electrostatic repulsion between
charged monomers. Therefore, the chain statistics is rod-like at
scales larger than the blob $\xi$ (in the plane parallel to the
surface) and Gaussian at smaller scales  ($\xi\simeq ag^{1/2}$).
However, the blob size is found by assuming that the electrostatic
attraction to the surface is of order $k_BT$:
$\frac{fg\sigma_{0}\xi^{2}e^{2}}{\epsilon \xi}\sim k_BT$ which
leads to
$\xi\simeq\xi_e\left(\frac{\sigma_c}{\sigma_0}\right)^{1/3}>\xi_e$.
The chain segment is thus confined by the electrostatic attraction
to the charged surface. However, contrary to the case of dilute
solutions (single chain adsorption), polyelectrolytes do not lie
flat on the surface since, as mentioned in the Introduction, the
system gains conformational entropy by allowing more chains to
adsorb with large loops protruding in the solution. The
concentration of blobs in this region is found by assuming that
this compensation layer actually compensates the surface charges
and that the thickness is given by the confinement blob size
$\xi$: \begin{equation}
\Lambda_1\simeq\xi_e\left(\frac{\sigma_c}{\sigma_0}\right)^{1/3}
\label{lambda1inf}
\end{equation}
This scaling result has been shown numerically for semi-dilute solutions by
Borukhov \textit{et al.}~\cite{Borukhov} and Shafir \textit{et
al.}~\cite{Shafir}  using the mean-field equations. The electric field vanishes
at $z=\Lambda_1$ in this regime. Of course, we do not have a closed packing of
these blobs, $\xi$, since the number of blobs per unit surface is
$\sigma_0/fg\simeq(\sigma_0/\sigma_c)^{5/3}\xi_e^{-2}<\xi_e^{-2}$.

This picture is valid only if the average monomer volume fraction close to
the surface is higher than the bulk value, i.e. $\phi_e(\sigma_0/
\sigma_c)^{4/3}>\phi_b$ or
\begin{equation}
\sigma_c\left(\frac{\phi_b}{\phi_e}\right)^{3/4}\leq\sigma_0\leq\sigma_c
\end{equation}

\noindent{\bf III.A.2 Case $\sigma_{0}>\sigma_{c}$.}

In the limit $\sigma_{0}\gg\sigma_{c}$, the monomer volume fraction close to
the surface, $\phi_0$, increases being larger than $\phi_e$, the
electrostatic attraction becomes larger than the electrostatic
intra-repulsion, and the osmotic pressure tends to swell the compensation
region. Assuming that excluded volume interactions dominate over elasticity,
as in Ref.~\cite{DobryninDeshkovski}, solving eq.~(\ref{PBE4}) of Appendix~2
yields (at the leading order in $\phi_{e}/\phi_{0}\ll 1$ and assuming $\mu\simeq
g_e^{-1}$ in this compensation region) a parabolic profile:
$\phi(z)\cong\phi_e+(\phi_0-\phi_e)(1-z/\Lambda_1)^2$, where
$\phi_0=(3\pi\ell_B\sigma_0^2)^{1/3}$ and
\begin{equation}
\Lambda_{1}\cong\xi_{e}\left(\frac{\sigma_{0}}{\sigma_{c}}\right)^{1/3}
\label{lambda1sup}
\end{equation}
The structure of the compensation region is thus self-similar (see
Fig.~\ref{fig3}) as already described for the dilute case in
Ref.~\cite{DobryninDeshkovski}. For the general case, the polymer
elasticity is no more negligible and the full
equation~(\ref{PBE4}) must be solved without any approximation.

It is important to note that our result Eq.~(\ref{lambda1sup}) is in
contradiction with Andelman \textit{et al.}~\cite{Borukhov,Andelman,Shafir}.
They predict that the characteristic length scale of the adsorbed layer
decreases with $\sigma_{0}$~(\cite{Andelman}, eq.~(71)) following
Eq.~(\ref{lambda1inf}). However, we argue that, for strong adsorption
($\sigma_{0}\gg\sigma_{c}$), the structure of the compensation region is found
by balancing the electrostatic adsorption with the osmotic pressure (excluded
volume interactions) and that chains are not confined. Instead of blobs of
confinement, the structure is described in terms of osmotic blobs, in a similar
way of describing neutral polymer layers.

Anyway, as we argue below (Section~4), in the semi-dilute regime and for long
polyelectrolytes, we do not expect that this region has a dominant influence on
observables such as the thickness of the layer, or the amount of material
attached to the surface, and it is enough for our purpose to treat the
restricted case where $\sigma_0\approx \sigma_c$, and
\begin{equation}
\Lambda_1 \cong \xi_e  \label{lambda1}
\end{equation}
Anticipating the use of the SFT, it is useful to estimate the amount of
monomers (per tail), $n_1$, involved in the compensation region. In a
simple view where each tail contributes to one electrostatic blob, we find
\begin{equation}
n_1\cong g_e  \label{n1}
\end{equation}
and the local grafting density at the outer limit of the compensation region
scales as
\begin{equation}
S(n_1)\cong \frac1{\xi_e^2}  \label{S(n1)}
\end{equation}

\subsection{Intermediate region (2).}

In this region, the direct electrostatic influence of the surface vanishes,
and from this point of view, the structure should be similar to a layer of
polyelectrolytes chains ``adsorbed'' onto a fictive neutral surface
positioned at $z=\Lambda_1$. Here, the concentration of monomers decreases
from $\phi_e$ to the bulk value, $\phi_b$. Because monomers and
counter-ions are ruled by very different equations, the electrostatic
potential $\Psi(z)$ does not vanish. As the system is globally neutral and
$\phi(\Lambda_1)=\phi_e>\phi_b$ in the semi-dilute regime, charge conservation in the system implies that the volume fraction of counter-ions, $\rho^-(z)$, will decrease from
$\rho^-(\Lambda_1)> f\phi_b$ to the bulk value $f\phi_b$ at the external border
of region~(2). Knowing that $\rho^-$ is related to the adimensional
electrostatic potential $\Psi=\beta eV$ by $\rho^-=f\phi_b\exp\Psi$ (see
Appendix~2), we deduce that $\Psi(\Lambda_1)=\Psi_1$ is positive (with the
boundary condition $\Psi(\infty)=0$). Thus, we expect that $\Psi(z)$ decreases
monotonically from a finite positive value, $\Psi_1$ (to be estimated below),
to 0.

\begin{figure}[htb]
\begin{center}
\includegraphics[height=5cm]{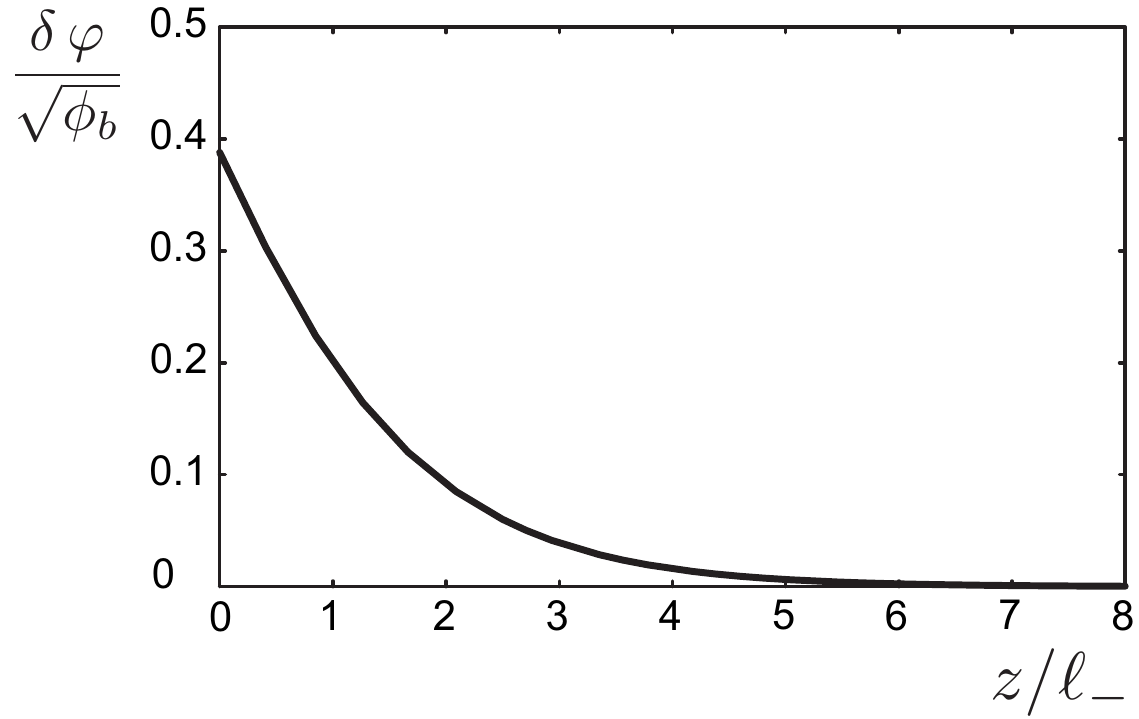}
\includegraphics[height=5cm]{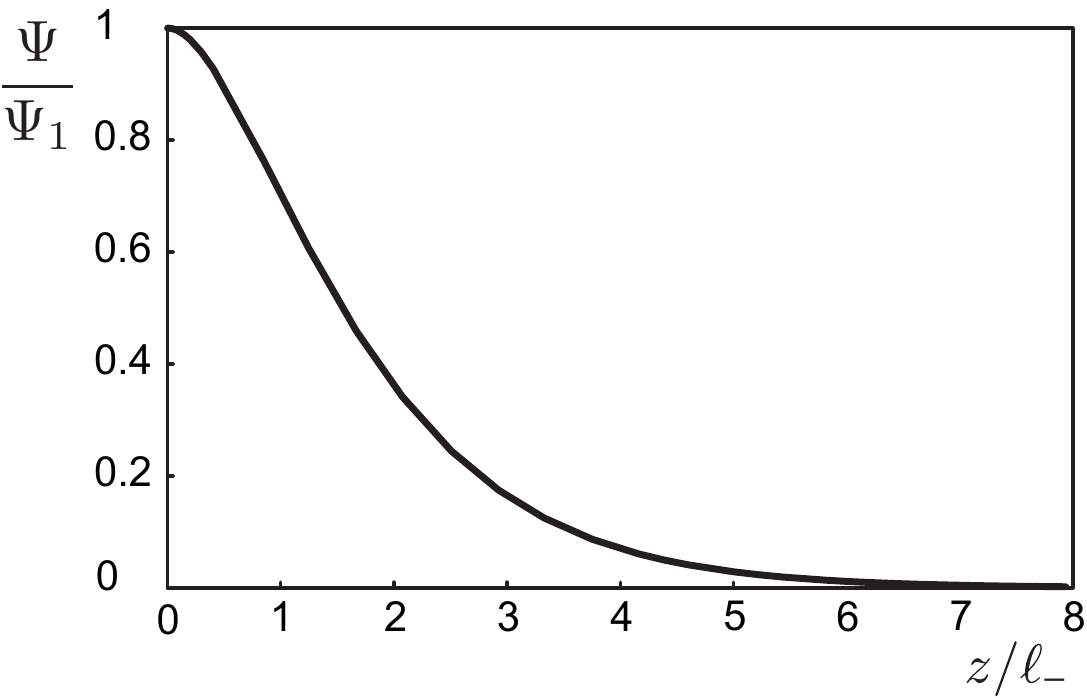}
\end{center}
\caption{Polymer order parameter (left), $\delta\varphi/\sqrt\phi_b$, and electrostatic potential (right), $\Psi/\Psi_1$, as a function of
the distance to $z=\Lambda_1$ (in units of $\ell_-$), in the intermediate region
for $\ell_B=1$, $f=0.1$, and $\phi_b=0.26$, in the mean-field approximation.}
\label{fig4}
\end{figure}

When $\phi_b$ is not too small in comparison to $\phi_{e}$, say
$\phi_{e}-\phi_{b}\ll\phi_{b}$, the linearized PBE system of equations described
in Appendix~2 is justified to find the relaxation of the concentration profile. Note that for the case $\phi_b\ll \phi_e$ the full system should be solved numerically~\cite{Borukhov,Shafir}. With the appropriate boundary conditions [$\phi(\Lambda_{1})=\phi_{e}$,
$\phi(\infty)=\phi_{b}$, $d\Psi/dz(\Lambda_{1})=0$ and $\Psi(\infty)=0$], we
find:
\begin{eqnarray}
\varphi(z) &=& \sqrt{\phi_{b}}+\left(\sqrt{\phi_{e}}-\sqrt{\phi_{b}}\right)
\left[(1-R)\,e^{-(z-\Lambda_{1})/\ell_{-}} + R\,e^{-(z-\Lambda_{1})/\ell_{+}} \right] \\
\Psi(z) &=& 8\pi\ell_{B}f\sqrt{\phi_{b}} \left(\sqrt{\phi_{e}}-\sqrt{\phi_{b}
}\right) \left[ \frac{1-R}{4\pi \ell_{B}f\phi_{b}-a^{2}\lambda_{-}} \right. \nonumber\\
&\times& \left. \mathrm{e}^{-(z-\Lambda_{1})/ \ell_{-}}
+ \frac{R}{4\pi\ell_{B}f\phi_{b}-a^{2} \lambda_{+}}
\mathrm{e}^{-(z-\Lambda_{1})/\ell_{+}}\right]  \label{psimf} \\
\rho(z) &=& f\phi_{b}[\Psi(z)+1]
\end{eqnarray}
where
\begin{eqnarray}
\ell_{\pm}^{-2} &=& \lambda_{\pm}=\frac{2\phi_{b}}{a^{2}}\left(\pi\ell_{B}f
+6\phi_{b}\right) [1\pm\sqrt{\Delta}]  \label{l+- mf} \\
\Delta &=&1- \frac{12\pi\ell_{B}f(2\phi_{b}^{2}+f)}{\phi_{b}(\pi\ell_{B}f+6
\phi_{b})^{2}} \\
R^{-1} &=&1-\sqrt{\frac{\lambda_{+}}{\lambda_{-}}}\frac{(4\pi \ell_{B}
f\phi_{b}-a^{2}\lambda_{-})}{(4\pi\ell_{B}f\phi_{b}-a^{2}\lambda_{+})}
\end{eqnarray}
In Fig.~\ref{fig4} are plotted $\delta\varphi(z)=\varphi(z)- \sqrt{\phi}_{b}$
and $\Psi(z)$ for $\ell_{B}=1$, $f=0.1$, and $\phi_b=0.26$ (concentrated semi-dilute solution). Note that the characteristic relaxation lengths $\ell_{\pm}$ were obtained by Ch\^{a}tellier and Joanny~\cite{Chatellier}. In this reference, the possibility of damped
oscillations in the volume fraction profile ($\ell_{\pm}$ imaginary) was
carefully examined for lower values of $\phi_b$. Here, we will only consider the situation
where $\ell_{\pm}$ are real numbers, for simplicity (the same discussion can be
carried on with the real parts of $\ell_{\pm}$ in more general cases). Hence the surface induce an external perturbation to $\varphi$ of magnitude $\sqrt{\phi_e}-\sqrt{\phi_b}$ which relaxes, at the linear response level, to the bulk value, $\sqrt{\phi_b}$ on the characteristic wave length of the fluctuations in the bulk.

Of course, we identify $\Lambda_2$ with the longest relaxation length,
$\ell_-$: 
\begin{equation}
\Lambda_2\cong \ell_- \label{lambda2}
\end{equation}
It is instructive to compare the characteristic relaxation length $\ell_{-}$
with $\xi_{b}$, which requires that we simplify the expression given by
eq.~(\ref{l+- mf}). Assuming that the intermediate region is dominated by the
counter-ions pressure with an effective excluded volume $v_{\rm eff}\sim f/\phi_b$ (and that contribution arising from the monomers is
negligible), the relaxation length $\ell_-$ writes (with $\Delta\simeq 1$):
\begin{equation}
\ell_-\approx \frac{a}{\sqrt{12\,f}}
\end{equation}
In this limit, we have $\ell_{-}\leq \xi_{b}$ for $\phi_{b}\leq
(f/\ell_{B})^{1/3}$, a reasonable case when $\phi_{b}<\phi_{e}$. The amount
of monomers is this region is given by
\begin{equation}
\Gamma_{2} = a^{-3}\int_{\Lambda_{1}}^{\Lambda_{1}+\Lambda_{2}}\phi (z)dz\label{gamma2} \cong
(0.17\phi_{b}+0.43\phi_{e}+0.40\sqrt{\phi_{b}\phi_{e}})\frac{\ell_{-}}{a^3}
\nonumber 
\end{equation}
Equation~(\ref{psimf}) also yields an estimate for the extremum value of the
electrostatic potential $\Psi_{1}=\Psi(\Lambda_{1})$:
\begin{equation}
\Psi_{1} \cong 8\pi \ell_{B}f\sqrt{\phi_{b}}\left(\sqrt{\phi_{e}}
-\sqrt{\phi_{b}}\right) \left[\frac{1-R}{4\pi\ell_B f\phi_b-a^2\lambda_-}+\frac{R}{(4\pi\ell_B
f\phi_b-a^2\lambda_+)} \right]
\label{maxpotelec} 
\end{equation}
Equation~(\ref{maxpotelec}) in turn allows to estimate
\mbox{$\Psi_{0}=\Psi(z=0)$} through the integrated Poisson equation:
\begin{equation}
\Psi_{0}=\Psi_{1}-4\pi\ell_{B}
\sigma_{0}\Lambda_{1}+4\pi\ell_B f\int_{0}^{\Lambda_{1}}\int_{0}^{u}\phi(z)
dzdu
\end{equation}
For the moderate adsorption case ($\sigma_{0}\approx\sigma_{c}$) that we
consider in this article, we find:
\begin{equation}
\Psi_0\cong \Psi_1 -2\pi \ell_{B}\frac{fg_e}{\xi_e}
\end{equation}
For the case of strong adsorption ($\sigma_{0}\gg\sigma_{c}$), we obtain:
\begin{equation}
\Psi_0\cong \Psi_1 -2\pi \ell_B\frac{fg_e}{\xi_e}
\left(\frac{\sigma_0}{\sigma_c}\right)^{4/3}
\end{equation}
The SFT approach described in Section~2 is unlikely to describe correctly
the intermediate region. At a simple scaling level, we will note that the
local grafting density should drop from $S(n_{1})\cong \frac{1}{\xi_{e}^{2}}$
to $S(n_{2})\cong \frac{1}{\xi_{b}^{2}}$, where $n_{2}$ is the curvilinear
index at $z=\Lambda_1+\Lambda_2$. A crude estimate for $n_{2}$ is found with
a renormalized linear string approximation for the tails in the intermediate
regime:
\begin{equation}
n_{2}\cong n_1+g_e\frac{\Lambda_2}{\xi_e}\simeq g_{b}  \label{n2}
\end{equation}
which is equivalent to write
\begin{equation}
\Lambda_2\cong \xi_b-\xi_e\simeq \xi_b  \label{L2}
\end{equation}
where the last approximation holds in the limit $\phi_e\ll\phi_b$. Hence,
the thickness of region~(2) is then given by equation~(\ref{lambda2}) at a
mean-field level and for $\phi_{e}-\phi_{b}\ll\phi_{b}$, and by
equation~(\ref{L2}) at a scaling level.

\subsection{Outer region (3).}

Here, the polymer concentration (resp. the counterion concentration, the
adimensional electrostatic potential) is the bulk concentration, $\phi_{b}$
(resp. $f\phi_{b}$, 0 at scales larger than the bulk blob size), but the tails
and loops in this region belong to chains in direct contact with the solid
surface. At a scaling level of description, the layer is a close packing of
semi-dilute blobs (size $\xi $). By analogy with the Rubinstein description of
the bulk solution~\cite{Rubinstein}, we expect that electrostatic interactions
are screened at distances larger than the mesh size $\xi$, and the system is
amenable to a description in terms of neutral chains. Hence the SFT approach put
forward in Section~2 is now appropriate.

\begin{figure}[htb]
\begin{center}
\includegraphics[height=7cm]{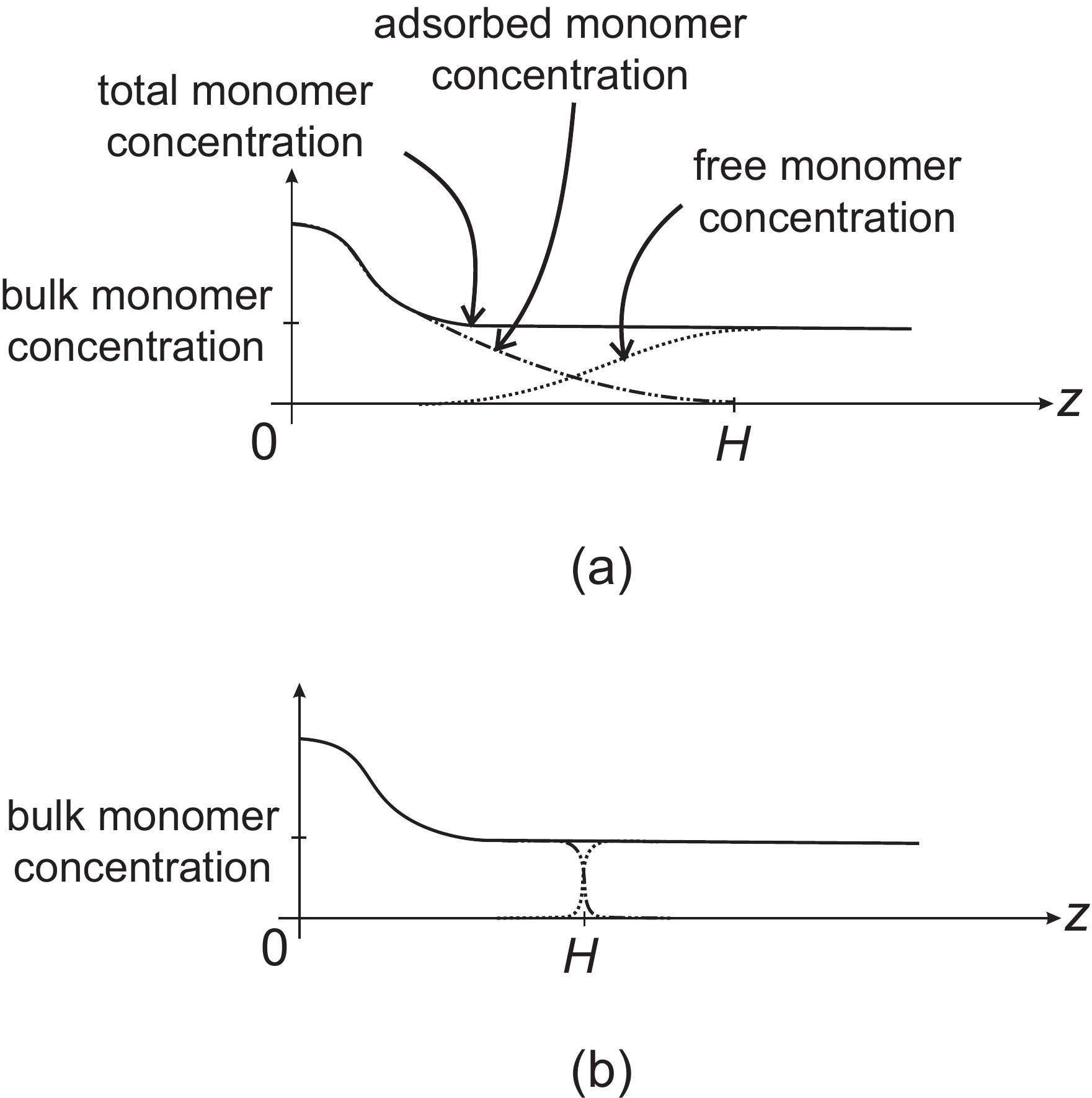}
\end{center}
\caption{We expect a slowly vanishing concentration profile for
the monomers which belong to chains in direct contact with the surface (a).  In
our approach, we simplify this picture into a sharp, Heaviside type of profile
at the outer fringe (b).}
\label{fig5}
\end{figure}

Obviously, free and adsorbed chains interpenetrate, and we expect
a slowly vanishing concentration profile for the monomers which
belong to chains in direct contact with the surface, as depicted
in Fig.~\ref{fig5}a. In what follows, however, we assume that the free
chains do not penetrate in the layer of tails and loops in direct
contact with the surface. Accordingly, we consider a sharp,
Heaviside type of profile at the outer fringe (see Fig.~\ref{fig5}b). A
similar type of simplification is found in various theories of
polymeric layers. In the context of polymer brushes in good
solvent, \textit{e. g.}, Alexander and de Gennes assume an
Heaviside step for the concentration profile in order to find the
extension of the layer~\cite{Alexander,PGGbrush}. Their scaling
result was confirmed by more refined theories where this
constraint is relaxed (instead a parabolic profile is
found)~\cite{MWC}. Similarly, Guiselin assumes no interpenetration
in the context of molten solutions of neutral polymers adsorbed at
an interface~\cite{Guiselin}. Here again, the scaling result was
experimentally confirmed by measuring the amount of material
irreversibly adsorbed~\cite{Liliane}. Because the semi-dilute
solution is amenable to a description in terms of neutral blobs of
strings of electrostatic coils~\cite{PGGPincus,Rubinstein}, it is
likely that the situation is analogous. In any case, we are
confident that the \textit{scaling results} that we find will not
be dramatically affected by our simplification.

\begin{figure}[htb]
\begin{center}
\includegraphics[height=5cm]{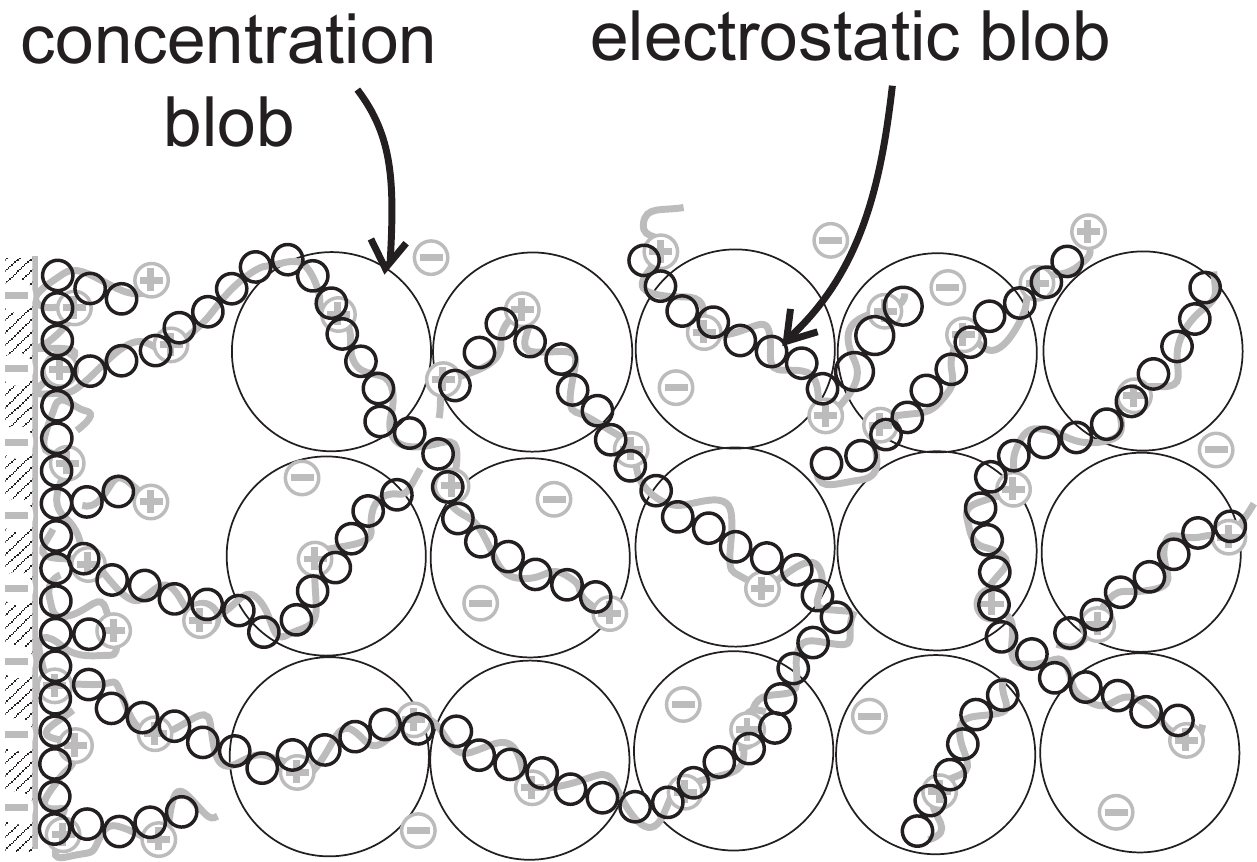}
\end{center}
\caption{Scaling representation of the adsorbed layer of
polyelectrolytes in the case $\sigma_{0}\simeq\sigma_{c}$ in terms of
electrostatic blobs (size $\xi_{e}$) and concentration blobs (size $\xi_{b}$).}
\label{fig6}
\end{figure}

The SFT equations suitable to describe this region
are~(\ref{SFT1})--(\ref{SFT2}). In equilibrium, we expect that the bulk osmotic
pressure exerted at the extremity of the layer ($z=H$) is balanced by the inner
osmotic pressure, and the elasticity can be neglected in eq.~(\ref{SFT2}). This
yields
\begin{equation}
\frac{S}{\dot{z}}\cong \phi _{b}  \label{s/zpoint}
\end{equation}
and thus [eq.~(\ref{phi})], $\phi(z)\cong \phi_{b}$. When this relation is
introduced into eq.~(\ref{SFT1}), this equation simplifies into
\begin{equation}
\frac{\xi_{e}^{1/2}}{\phi_{b}^{3/2}}S^{2}+\frac{S''}{S'} \cong 0
\end{equation}
which admits a power law type of solution in the limit $N\rightarrow\infty$,
\begin{equation}
S(n)\cong \frac{\phi_{b}^{3/4}}{a^{2}(\xi_{e}/a)^{1/4}n^{1/2}}
\label{Souter}
\end{equation}
Note that we find $S(n_{2})\cong \xi_b^{-2}$. Now solving
Eqs.~(\ref{s/zpoint})--(\ref{Souter}) (with the boundary condition
$z(n_2)\cong \Lambda_1+\Lambda_2$) for $z$, gives a Gaussian structure
at large scale: \begin{equation}
z(n)\cong \Lambda_{1}+\Lambda_{2}+a(\xi_{e}/a)^{-1/4}\phi_{b}^{-1/4}n^{1/2}
\end{equation}
which readily yields [$H=z(N)$]:
\begin{equation}
\Lambda_3 \cong a\frac{N^{1/2}}{(\xi_e/a)^{1/4}\phi_b^{1/4}}  \label{lambda3}
\end{equation}
and
\begin{equation}
\Gamma_3 = \int_{n_2}^N S(n)\, dn \cong
\frac{\phi_b^{3/4}N^{1/2}}{a^2(\xi_e/a)^{1/4}}  \label{gamma3}
\end{equation}
Note that the result eqs.~(\ref{lambda3})--(\ref{gamma3}) are precisely what
we would find with scaling arguments assuming that the chains are Gaussian
at large scale, and the SFT is only a formal way to recover them in this
context. Figure~\ref{fig6} shows a scaling representation of the whole adsorbed
layer in the case $\sigma_0\simeq\sigma_c$. On the other hand, $S$ is directly
related to the loop size distribution $P$ through~\cite{ManoPRE}
\begin{equation}
S(n)=\int_n^N\,P(u)\,du
\end{equation}
and eq.~(\ref{Souter}) may be experimentally checked with a surface force
apparatus~\cite{senden}.

\section{Discussion}


The structure that we propose for the adsorbed layer clearly separates the
influence of the charged surface which occurs only in the compensation region, from that of the bulk solution which act on the outer region, with an intermediate regime in between. We briefly describe the influence of $\phi_b$ and $\sigma_0$ on the layer structure and then discuss the occurrence of charge inversion for layers of very long charged chains.

We expect that as we increase $\phi_b$ from $\phi_b^*$, we progressively build the
outer region at the expense of the intermediate region ($\Lambda_2$ decreases), but without much consequences for the compensation region. This is because the osmotic
pressure of the counter-ions (the dominant contribution to the physics in the
intermediate region) is much higher near $z=\Lambda_1$ than at
$z=\Lambda_2$. Arguably, any increase of the bulk concentration of counter-ions
will predominantly affect the low osmotic domain of the intermediate region, and
hardly the structure near $z=\Lambda_1$, where the pressure is high. Of
course, when $\phi_b$ reaches $\phi_e$, the intermediate region disappears
in our simple view, and the bulk solution should then affect directly the
compensation region. In this regime where fluctuations are negligible, numerical mean-field approaches are well adapted~\cite{Borukhov,wang}.
On the other hand, if we increase $\sigma_0$ from $\sigma_c$, at constant bulk concentration, the surface potential $|\Psi_0|$ increases and more and more monomers are involved
in compensating the surface charge ($n_1$ and $\Lambda_1$ increase).
Accordingly, the concentration profile in the compensating region is affected.
But for very long chains ($N\gg n_2$), we do not expect that this should
influence much the structure of the outer region.

Note that our description of the compensation and intermediate region
resembles, but is not quite similar to the description of the adsorbed layer
from dilute solution proposed in Ref.~\cite{DobryninDeshkovski}~: the
so-called self-similar adsorbed layer regime. Essentially, we keep the
parabolic decrease of the monomer volume fraction close to the surface,
until we reach $\phi=\phi_e$ (where neglecting the chain elasticity is no
more justified), and then we use the mean-field equations (which account for
the elastic contribution and the osmotic pressure of counter-ions) to find
the relaxation of $\phi$ below $\phi_e$. In Ref.~\cite{DobryninDeshkovski},
the region where $\phi\leq\phi_e$ is described in terms of electrostatic
blobs. However such a description does not take into account the presence of the
counter-ions in this region which is imposed by electroneutrality (the electric
potential is positive). In technical terms, we make advantage of the fact that
$\phi$ relaxes to a finite value, $\phi_b$, to determine the monomeric profile
below $\phi_e$ by linearizing the mean-field equations. Because $\phi$ goes to
$0$ in dilute solutions, such calculation is not possible and would necessitate
a numerical resolution. The difference between the two descriptions is not just
a matter of refinement since the relaxation of $\phi$ yields the variations of
the electrostatic potential. With our approach, we provide both a coherent and
complete picture for the variations of this quantity [eq.~(\ref{psimf})].\\

A crucial aspect of adsorbed polyelectrolyte layers is their ability to
capture a very large number of counter-ions per unit surface, much larger
indeed than $\ell_{B}^{-2}$. In particular, the driving force for the
adsorption of oppositely charged chains depends precisely on the amount of
\textit{potentially} free ions. With this in mind, it is interesting to
discuss the following situation.

Suppose that we build an adsorbed layer from a semi-dilute solution, as
described before. Then we remove the polymeric solution after the layer has
equilibrated, but we suppose that the adsorbed chains remain attached to the
charged surface, in a fashion similar to the scenario discussed by Guiselin
for irreversible adsorption of molten liquids of neutral chains~\cite{Guiselin}.
Finally, we put the decorated layer into a solution of pure solvent. The apparent charge (also called ''net charge" in Ref.~\cite{DobryninDeshkovski}) of the layer in the dilute solution will be
zero, because of the counter-ions present within (or in the proximity) of the
layer. On the other hand, the decorated layer carries a huge amount per unit
surface of \textit{potentially} free counter-ions. Provided that we disregard the
fluffy structure of the layer, the decorated surface effectively behaves as a
new charged surface of "nominal" charge density 
\begin{eqnarray}
\Delta\sigma \equiv \sigma -\sigma_0 &=& f(\Gamma_1+ \Gamma_2+\Gamma_3) -\sigma\nonumber \\
&=& f\Gamma_2 + \ell_B^{1/12} f^{7/6} \phi_b^{3/4} N^{1/2}
\end{eqnarray}
Hence, for long chains, we expect that $\Delta\sigma$ may be much larger than $\sigma_0$.

Let us compare: \textit{a)} a layer of polyelectrolytes reversibly
adsorbed from dilute solution in equilibrium, and \textit{b)} a
layer of polyelectrolytes irreversibly adsorbed from semi-dilute
solution put into a solution of pure solvent (the solid surfaces
have the same $\sigma_{0}$ in both cases). To this aim, we need a
precise vocabulary. We define three regimes, depending on the
comparison of $\sigma$ with $\sigma_{0}$ and $2\sigma_{0}$:
\begin{itemize}
\item $\sigma<\sigma_0$: undercharging ($\sigma =\sigma_0$ corresponding
to charge compensation);
\item $\sigma_0<\sigma<2\sigma_0$: overcharging ($\sigma =2\sigma_0$
corresponding to charge inversion);
\item $\sigma>2\sigma_0$: supercharging.
\end{itemize}
For case \textit{a)} the mean-field theory predicts exact charge
compensation in the absence of added salt~\cite{Joanny}. With added salt,
Joanny finds overcharging, with possibly charge inversion in the limiting
case of very high ionic force. In any case, there is no supercharging. It
also appears that the surface controls the amount of adsorbed material in
case \textit{a)} (together with the ionic strength of the solution in case
of added salt), not the chain length, nor the polymer concentration. A more
recent analysis is provided by the work of Dobrynin \textit{et
al.}~\cite{DobryninDeshkovski} who carefully combine the mean-field theory with
scaling arguments. But this study does not change the conclusions concerning the
regime of compensation.

In contrast, for case \textit{b)} our analysis strongly suggests that we
have supercharging for sufficiently long chains, $N>(\sigma_0/f-\Gamma_2)^2/[(\ell_B f^2)^{1/6}\phi_B^{3/2}]$, \textit{even in the absence of salt}.


\section{Concluding remarks}


In this paper, we describe completely the structure of the adsorbed layer from semi-dilute solution by identifying the physical mechanisms in each sublayers~: electrostatic attraction by the surface charge vs. three body interactions in the compensation region~(1), importance of the polymer elasticity and the counter-ions osmotic pressure in the intermediate region~(2), and presence of long tails and loops of adsorbed chains in the distal region~(3) at the bulk concentration. We give explicit expressions for the coverage including all monomers belonging to adsorbed polymers, $\Gamma$ and the layer thickness $H$ as a function of parameters, bare surface charge, bulk concentration which could be checked experimentally.

Clearly, our description of the intermediate region suffers from several
weaknesses. First, we consider a mean-field approach extrapolated outside
its domain of validity. Secondly, we use a linearized theory, which is only
justified for bulk volume fractions close to $\phi_e$. However, it is very
likely that our essential \textit{scaling type of conclusions} will be
robust to any improved description of the intermediate region.

Similarly, the influence of the solvent quality (unless it is bad solvent)
is rather limited. This is because the electrostatic interactions dominates
over the excluded volume interactions at large scales, and in actual facts,
the solvent only matters in the compensation region and in the electrostatic
blob. For example, it is simple to show that $\Gamma_3\cong
a^{-2}\phi_{b}^{3/4}(\ell_{B} f^{2})^{1/14}N^{1/2}$ and $\Lambda_3\cong
a\phi _{b}^{-1/4}(\ell_{B} f^{2})^{1/14}N^{1/2}$ in good solvent conditions.

This study is limited to the very low salt limit. When the ionic force is very large, electrostatic interactions are screened at length scale larger than the Debye length and lead to an effective excluded volume $v_{el}=f^2/(2n_s)$ where $n_s$ is the salt concentration~\cite{Joanny,Andelman}. The structure of the compensation is then modified but far from the surface, the theory of neutral polymer layer~\cite{Manomacromol} can be applied by introducing the total excluded volume $v+v_{el}$ (where $v$ is the bare one). One find for instance $\Gamma_3 \cong
a^{-2}\phi_b^{7/8}[v+f^{2}/(2n_s)]^{1/8}N^{1/2}$.

As far as we know, very few experimental measurements of the amount of material
attached to the surface as a function of $N$ have been done under the conditions
we are interested (oppositely charged surface, weakly charged chains, no
salt added, semi-dilute solutions). The experiments of Chibowski \textit{et al.}
on adsorption of polyacrylic acid (PAA) and polyacrilamide (PAM) to a
$\mathrm{Fe}_2\mathrm{O}_3$ solid surface, show a significant enhancement of
both the layer thickness and the amount of material when the chain molecular
weight is increased~\cite{Chibowski}. For instance, the thickness increases from
4.53 nm at molecular weight $M_w=$170 kg/mol to 6.69 nm at
$M_w=$240 kg/mol for PAA at pH 3 and for a polymer
concentration of 500 ppm (semi-dilute regime). This increase is in a qualitative
agreement with Eq.~(\ref{lambda3}) and is related to the formation of larger
loops and tails. Of course, the authors mention the important role of the pH,
since dissociation equilibriums are monitored by pH and thus both the number of
charges of the chains and of the substrate can vary. Such effect has to be
included in the theory to do a quantitative comparison.\\

Concerning the first step of the layer-by-layer process~\cite{Decher}, it should be pointed out that the adsorbed layer is globally neutral ($d\Psi/dz (H)=0$) because counter-ions which penetrate into the layer compensate the decorated surface charge. Hence it is not the apparent charge which is important but the amount of free counter-ions which can be released during the adsorption of oppositely charged polyelectrolytes. This is a key point  
since, by assuming that adsorbed chains stay upon rinsing, the driving force of adsorption is the exchange of counter-ions related to the amount of potentially free counter-ions. The good observable is thus the amount of adsorbed polymers $\Gamma$, and not the polymer surface excess relative to the bulk concentration. Hence the structure of the adsorbed layer, described in terms of loops and tails, is necessary to compute the coverage. We thus go beyond standard mean-field approaches~\cite{Joanny,varoqui,stoll1,stoll2,Borukhov,wang} which cannot distinguish between adsorbed chains and free chains in the distal region. As a result, we find that the issue of overcharging--supercharging relies more on the chain length $N$ and bulk volume fraction $\phi_b$ than salt concentration~: charge inversion occurs even at low salt concentration which underlines the role played by the chain length as a tunable experimental parameter. Presumably, this feature opens the way to a both simple and efficient way to tune the size of the successive layers in the Decher process. Obviously, a
closer look at the case where two layers of oppositely charged polymers
interpenetrate and rearrange~\cite{ladam} is required before we can drive any conclusion on the Decher process~\cite{castel}.\\
\\
We are grateful to L.~Bocquet, E.~Trizac, G.~Decher and R.R.~Netz for interesting discussions.

\appendix

\section*{Appendix 1 : scaling laws}


In this Appendix, we briefly recall the basic scaling concepts for
polyelectrolytes. We also provide the essential equations needed to
understand the scaling arguments presented in the article.
Our present understanding of weakly charged polyelectrolyte solutions
without added salt is based on the concept of electrostatic
blob~\cite{PGGPincus,Barrat}: because the different charges of the backbone
repel, the electrostatic interactions will deform the otherwise isotropic coil
into an elongated string. At a scaling level, the chain is pictured as a linear
string of electrostatic blobs (size $\xi_{e}$ , number of monomers $g_{e}$) such
that $\beta\frac{\left(fg_{e}e\right)^{2}}{\epsilon\xi_{e}}\cong 1$. Let us
assume $\Theta $ solvent conditions for simplicity (the two-body excluded volume
parameter $v=0$). Then the Flory scaling law for random walks ($\xi_{e}\cong
ag_{e}^{1/2}$) yields
\begin{equation}
\xi _{e}\cong \frac{a}{\left(f^{2}\ell_{B}\right)^{1/3}}\qquad \mathrm{and}
\qquad g_{e}\cong \frac{1}{\left(f^{2}\ell_{B}\right)^{2/3}}
\end{equation}
where $\ell_{B}=\frac{\beta e^{2}}{4\pi\epsilon a}$ is the Bjerrum length, in
units of $a$. The extension of the polymer is
\begin{equation}
L\cong \frac{N}{g_{e}}\xi_{e}\cong aN\left(f^{2}\ell_{B}\right)^{1/3}
\end{equation}
and the local volume fraction (inside the electrostatic blob),
\begin{equation}
\phi_{e}\cong \left(f^{2}\ell_{B}\right)^{1/3}  \label{phie}
\end{equation}
When the bulk concentration of monomers is larger than the overlapping
concentration, $\phi_{b}^{*}\cong NL^{-3}\sim N^{-2}$, the widely accepted
picture is that electrostatic interactions are screened at distances larger
than the mesh size of the solution~\cite{Rubinstein}. Accordingly, the
solution is a close packing of semi-dilute blobs (size $\xi_{b}$, number of
monomers $g_{b}$, $\phi_{b}\cong g_{b}/\xi_{b}^{3}$), such that the chain is
a Gaussian walk of blobs at a scale larger than $\xi_{b}$, and a linear
string of electrostatic blobs at a scale smaller than $\xi_{b}$. We
have~\cite{PGGPincus}
\begin{equation}
\xi_b \cong a\frac{\phi_b^{-1/2}}{\left(f^2\ell_B\right)^{1/6}} \qquad
\mathrm{and} \qquad g_b \cong \frac{\phi_b^{-1/2}}{\left(f^2\ell_B
\right)^{1/2}}
\end{equation}
Eventually, when the bulk volume fraction exceeds $\phi_{e}$, the
electrostatic blob vanishes, and the chain is Gaussian at all scales.


\section*{Appendix 2 : mean-field theory}


In this Appendix, we recall the equations of the mean-field theory used to
describe polyelectrolytes (see, e.g.~\cite{Chatellier,Andelman}). The basic
idea is to describe the monomers as microscopic ions carrying a charge $fe$,
with the following specific features: \textit{a)} the translational entropy
is negligible (this contribution scales as $(\phi /N)\log (\phi /N)$ and
vanishes for long polymers ($N\gg 1$)), \textit{b)} an excluded volume
interaction estimated in mean-field, \textit{c)} an elastic contribution to
account for the chain deformation. Then the energy (per cm$^2$) writes
\begin{eqnarray}
\beta F &=&-\frac{1}{2a^2}\sigma_0\Psi(0)
+ \int_{0}^{\infty} \left[ \Psi\rho^{+}-\Psi
\rho^{-}-\frac{a^{2}}{8\pi\ell_{B}} (\nabla\Psi)^{2}\right]
\frac{dz}{a^3}\label{Mean-Field energy}\\ &+& \int_{0}^{\infty}
\left[\rho^{+}(\ln\rho^{+}-1) +
\rho^{-}(\ln\rho^{-}-1)\right]\frac{dz}{a^3} +
\int_{0}^{\infty}\left[\frac{1}{3}\phi^{3}+\frac{a^{2}}{12}
(\nabla\sqrt{\phi})^{2}+f\phi\Psi\right] \frac{dz}{a^3} \nonumber
\end{eqnarray}
where the first term is the self-energy of the surface, the second term is
the electrostatic energy of resp. the co-ions and the counter-ions and the
self-energy of the electric field, the third term is the translational
entropy of micro-ions (in volume fraction $\rho^{\pm}$), and the fourth term
accounts for the energy of the monomers (resp.: three body-interaction with
amplitude $w=a^{6}$, Edwards elastic contribution, electrostatic energy of the
charged monomers). Minimizing the energy yields the Poisson-Boltzmann-Edwards (PBE) system of equations:
\begin{eqnarray}
\frac{d^2\Psi}{dz^2}(z) &=& -\frac{4\pi\ell_B}{a^2}\left[f\phi(z)
+n_s\mathrm{e}^{-\Psi(z)} -(n_s+f\phi_b)
\mathrm{e}^{\Psi(z)}\right] \label{PBE1} \\ 
\rho^+(z) &=& n_s\mathrm{e}^{-\Psi(z)}
\label{PBE2} \\ \rho^-(z) &=& (n_s+f\phi_b) \mathrm{e}^{\Psi(z)} \label{PBE3}\\
\frac{a^2}{6}\frac{1}{\sqrt{\phi(z)}}\frac{d^2\sqrt{\phi(z)}}{dz^2}
&=& \phi(z)^2+ f\Psi(z)-\mu \label{PBE4}
\end{eqnarray}
where $n_s$ is the volume fraction of salt in the bulk.
Note that the mean-field theory is strictly valid for concentrations above
$\phi_{e}$ for at least one reason: the average volume fraction, $\phi $, is not
the local volume fraction experienced by the monomers, $\phi_{e}$, in the
semi-dilute regime, and the three-body interactions (the term $\phi^{3}$ in
eq.~(\ref{Mean-Field energy})) is therefore incorrect. However, in the absence
of any other theory to describe these systems, it is quite common to extrapolate
the mean-field approach outside its strict domain of validity, and try to solve
the PBE system of equation in the semi-dilute regime.
Because $\mu$ is the chemical potential of the monomers in the solution,
and this quantity is sensitive to the local concentration, it is tempting to
correct for $\mu \cong 1/g_e$ ($\phi_e^2$ in theta solvent),
while keeping the mean-field estimate of the chemical potential ($\phi(z)^2$
in our context) in the rhs of eq.~(\ref{PBE4}). However, this approximation has
a technical drawback since $\Psi(z)=0$, $\phi(z)=\phi_b$ is then not solution of
the Edwards equation, as we expect far away from the surface. Here, we will keep
the PBE system of equation with $\mu\cong\phi_b^2$.
In Section~3.2, we use the Ch\^{a}tellier and Joanny
approximation~\cite{Chatellier} where both the Boltzmann
eq.~(\ref{PBE2}--\ref{PBE3}) and Edwards eq.~(\ref{PBE4}) are linearized (resp.
around $\Psi =0$, and $\phi=\phi_{b}$). In terms of the polymer order
parameter $\varphi=\sqrt{\phi} $, the PBE system of equation simplifies into
\begin{eqnarray}
\frac{d^2\Psi}{dz^2}(z) &=& -\frac{8\pi\ell_{B}f}{a^2} 
\sqrt{\phi_{b}}\,\delta \varphi(z) + \left(\kappa^2 +
\frac{4\pi\ell_B}{a^2}\right)\Psi(z)  \label{LPBE1} \\ \rho^+(z) &=& n_{s}[1-\Psi (z)]
\label{LPBE2} \\ \rho ^{-}(z) &=& \left(n_{s}+f\phi _{b}\right) [1+\Psi (z)]
\label{LPBE3} \\
\frac{a^{2}}{6\sqrt{\phi_{b}}}\frac{d^{2}\delta\varphi}{dz^{2}}(z) &=&
4\phi_{b}^{3/2}\delta\varphi(z)+f\Psi(z)  \label{LPBE4} \end{eqnarray}
where $\Psi\ll 1$, $\varphi=\sqrt{\phi_{b}}+\delta\varphi$ (with $\delta
\phi\ll\sqrt{\phi_{b}}$) and \mbox{$\kappa^{2}=8\pi\ell_B n_s/a^2$}.

\end{document}